\documentclass[technote,10pt]{IEEEtran}
\usepackage{amsmath,amsfonts}
\usepackage{algorithmic}
\usepackage{array}
\usepackage[caption=false,font=footnotesize,labelfont=rm,textfont=rm]{subfig}
\usepackage{color} 
\usepackage{diagbox}
\usepackage{textcomp}
\usepackage{stfloats}
\usepackage{url}
\usepackage{verbatim}
\usepackage{graphicx}
\def\BibTeX{{\rm B\kern-.05em{\sc i\kern-.025em b}\kern-.08em
		T\kern-.1667em\lower.7ex\hbox{E}\kern-.125emX}}
\usepackage{balance}
\usepackage{hyperref}
\usepackage{makecell}
\usepackage{threeparttable}
\usepackage{multirow}
\usepackage{algorithm}
\usepackage{caption}
\usepackage{cite}
\usepackage{setspace}
\usepackage{bm}
\usepackage{url}
\usepackage{numprint}
\usepackage{siunitx}
\usepackage[font=footnotesize]{caption}
\usepackage{flushend}
\begin{document}
	
\begin{spacing}{0.95}	
		
\title{CSI-GPT: Integrating Generative Pre-Trained Transformer with Federated-Tuning to Acquire Downlink Massive MIMO Channels}
		
\author{Ye Zeng, Li Qiao,  Zhen Gao, Tong Qin, Zhonghuai Wu, Emad Khalaf, Sheng Chen,~\IEEEmembership{Life Fellow,~IEEE}, and Mohsen Guizani,~\IEEEmembership{Fellow,~IEEE}

\vspace*{-3.0mm}
\thanks{Ye Zeng, Li Qiao,  Zhen Gao, Tong Qin, and Zhonghuai Wu are with the School of Information and Electronics, Beijing Institute of Technology, Beijing 100081, China (e-mail: {gaozhen16}@bit.edu.cn).}
\thanks{Emad Khalaf is with Electrical and Computer Engineering Department, Faculty of Engineering, King Abdulaziz University, Jeddah 21589, Saudi Arabia (E-mail: ekhalaf@kau.edu.sa).}  %

\thanks{S. Chen is with the School of Electronics and Computer Science, University
	of Southampton, Southampton SO17 1BJ, U.K., and also with King Abdulaziz
	University, Jeddah 21589, Saudi Arabia (e-mail: sqc@ecs.soton.ac.uk).}  %
\thanks{Mohsen Guizani is with Machine Learning Department, Mohamed bin Zayed University of Artificial Intelligence (MBZUAI), UAE (e-mail: mguizani@ieee.org).}  %
}
\maketitle
\vspace{-2mm}
\begin{abstract}
In massive multiple-input multiple-output (MIMO) systems, how to reliably acquire downlink  channel state information (CSI)  with low overhead is challenging.
			In this work, by integrating the  generative pre-trained Transformer (GPT) with federated-tuning, we propose a CSI-GPT approach to realize efficient downlink CSI acquisition.
			Specifically, we first propose a Swin Transformer-based channel acquisition network (SWTCAN) to acquire downlink CSI, where pilot signals, downlink channel estimation, and uplink CSI feedback are jointly designed. Furthermore, to solve the problem of insufficient training data, we propose a variational auto-encoder-based channel sample generator (VAE-CSG), which can generate sufficient CSI samples based on a limited number of high-quality CSI data obtained from the current cell. 
			The CSI dataset generated from VAE-CSG will be used for pre-training SWTCAN.    
			To fine-tune the pre-trained SWTCAN  for improved performance,  we propose an online  federated-tuning method,  where  only a small amount of SWTCAN parameters are unfrozen and updated using over-the-air computation, avoiding the high communication overhead caused by aggregating  the complete CSI samples from user equipment (UEs) to the BS for centralized fine-tuning.
			Simulation results verify  the advantages of the proposed SWTCAN and the communication efficiency of the proposed federated-tuning method.  Our code is  publicly available at \url{https://github.com/BIT-ZY/CSI-GPT} \color{black}
\end{abstract}

\vspace{-2mm}		
	
\vspace{-2mm}	
\section{Introduction}\label{S1}

In massive multiple-input multiple-output (MIMO) systems, accurate downlink channel state information (CSI) is crucial for beamforming and resource allocation. However, in frequency division duplexing (FDD) systems, accurate estimation and feedback of downlink CSI with low pilot/feedback overhead is challenging, due to the high-dimensional CSI caused by the large number of antennas at the base station (BS) and the non-reciprocity between uplink and downlink channels \cite{liu2024,gao2015spatially,jiang2024,zhang2024}. 
		
As the channel gains associated with different antennas are correlated, the massive MIMO CSI matrix has the inherent redundancy, which can be exploited to reduce the CSI acquisition overhead \cite{Xu2024,Zhu2023,Fang2024,Ye2024}. 
Due to its powerful capabilities of feature perception and extraction, deep learning (DL) has been widely used to process CSI in massive MIMO systems for various tasks. The authors of \cite{mxs_tvt} proposed a DL-based joint pilot design and channel estimation scheme, where the fully connected (FC) layer and the convolutional neural network (CNN) are used to design the pilot signal and estimate the CSI, respectively. The authors of \cite{Mahdi_TWC_estimation} made improvements to \cite{mxs_tvt} by designing pilot signals on different subcarriers differently to further reduce the pilot overhead. As for CSI feedback, the seminal work \cite{csinet} proposed an autoencoder (AE)-based end-to-end (E2E) optimization framework, and the  authors of \cite{csinet_plus} improved CSI feedback performance by introducing  receptive fields of different sizes for CSI feature extraction.

To improve the deployability of CSI feedback, the authors in \cite{Nerini2023} proposed a scheme in which the length of the feedback codeword is variable with the sparsity of the channel. The authors in \cite{guo2022} proposed a scheme based on knowledge distillation. Both schemes largely reduce the complexity of the network that needs to be deployed. In addition, the latest information about the application of DL in CSI feedback and the comparison can be obtained from \cite{guo2022overview}.
\color{black}		

More recently, as reported in \cite{Jiajia_estimation_feedback,guo2021canet}, joint design of pilot signals, downlink CE and CSI feedback using DL can further improve the downlink CSI acquisition  performance. Another approach to improve the performance is to exploit more advanced neural network (NN) architectures. The authors of \cite{wangyang_magzine} investigated the application of Transfomer architecture in various massive MIMO processing tasks, which consistently shows advantages over CNN-based algorithms. With the development of NN architectures, variants of Transformer, e.g., Swin Transformer, have shown better feature capture capability in image processing tasks \cite{Transformer_survey}, which is another blessing of DL-based massive MIMO signal processing.		
		
The current DL models \cite{mxs_tvt,Mahdi_TWC_estimation,csinet,Jiajia_estimation_feedback,csinet_plus,guo2021canet,wangyang_magzine} need a large amount of high-quality CSI samples for training, which can be obtained from actual measurements or generated from classical channel models, e.g., COST 2100 and clustered delay line (CDL) channel \cite{3gpptr38901}. However, it is either difficult or communication-inefficient to obtain a large number of actual CSI samples in various practical scenarios \cite{channelGan}, and it may degrade the NN performance if the features of training dataset and test dataset are not consistent. To overcome this issue, the generative adversarial network (GAN) is employed in \cite{channelGan} to generate CSI training datasets based on only a small amount of CSI measurements. Another promising solution \cite{yang2022federated} is to use federated learning (FL) to avoid collecting high-dimensional actual CSI samples, thereby reducing the communication overhead considerably. By exploiting gradient compression and over-the-air computation (AirComp), the authors of \cite{amiri2020federated} proposed a communication-efficient FL framework for image classification tasks. 
The authors further proposed a massive digital AirComp scheme that are compatible with the current wireless networks in \cite{qiao2024}.
\color{black}
To tune a large NN model, e.g., Transformer, in a communication-efficient manner, the authors of \cite{chen2022fedtune} utilized FL to fine-tune part of the parameters of a pre-trained Transformer model. The authors of \cite{Ahmet_fl_estimation_risMIMO, DLL_FL_estimation_ris, FL_estimation_feedback} showed the potential of FL-based CE and/or CSI feedback for massive MIMO systems. They also showed that users' data privacy can be protected by using FL. However, whether FL is more communication-efficient than conventional centralized learning (CL) that requires the feedback of CSI samples from user equipment (UEs) to BS remains unexplored.
		
\begin{figure}[t]
\begin{center}
\includegraphics[width=\columnwidth]{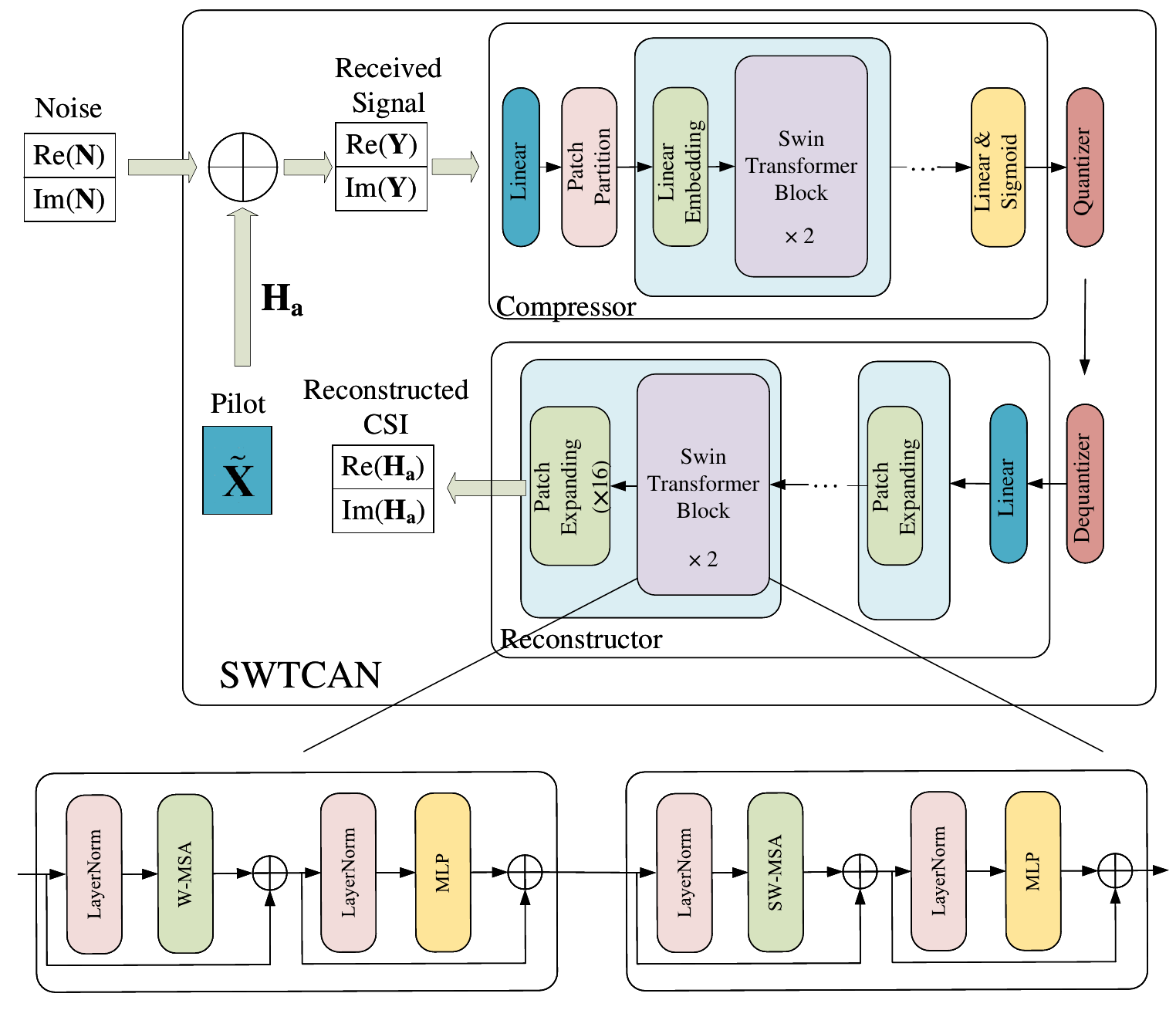}
\end{center}
\vspace*{-4.5mm}
\caption{Structure of the proposed SWTCAN.\color{black}}
\vspace{-7mm}
\label{fig1}
\end{figure}

In this paper, by integrating the  generative pre-trained Transformer (GPT) with federated-tuning, we propose a CSI-GPT approach to realize efficient downlink CSI acquisition. Our main contributions can be summarized as follows.
\begin{itemize}
\item {We propose a Swin Transformer-based channel acquisition network (SWTCAN) as shown in   Fig. \ref{fig1} to acquire downlink CSI with lower pilot/feedback overhead, where downlink pilot signal, CE and CSI feedback are jointly designed. Our SWTCAN not only retains the extraction capability of the conventional Transformer-based approach \cite{wangyang_magzine} but also overcomes its weakness in multi-scale feature extraction. Consequently, lower pilot and feedback overhead can be achieved. 
}
\setlength{\itemsep}{3pt}
\item {We propose a variational AE-based channel sample generator (VAE-CSG), which can effectively solve the problem of insufficient high-quality CSI samples. Since channel features in different cells vary dramatically, to maximize the potential of SWTCAN, its training at different BSs should rely on large numbers of CSI samples of the respective cells. However, the number of high-quality CSI samples from the current cell is usually limited. We propose a pre-trained strategy by pre-training VAE-CSG using a large number of CSI samples that typically have different features from those of the current cell. Subsequently, we fine-tune VAE-CSG using a limited number of high-quality CSI samples from the current cell. The fine-tuned VAE-CSG then generates a large number of CSI samples for pre-training SWTCAN.}
\setlength{\itemsep}{3pt}
\item {Finally, to fine-tune the pre-trained SWTCAN for improved performance, we propose an online federated-tuning method. Only a small amount of SWTCAN parameters (around 11\%) are unfrozen and updated using AirComp, avoiding the high communication overhead caused by aggregating the CSI samples from UEs to the BS for centralized fine-tuning.The simulation results demonstrate that in typical cases, the proposed federated-tuning method can reduce uplink communication overhead by up to 34.8\% compared to the traditional CL method.}
\end{itemize}
\vspace{-1mm}
\textit {Notation}: Boldface lower and upper-case symbols denote column vectors and matrices, respectively. 
Superscripts $(\cdot)^{\rm T}$ and $(\cdot)^{\rm H}$ denote the transpose and conjugate transpose operators, respectively.
$p( \mathbf{x\mid y})$ is the conditional distribution of $\mathbf{x}$ given $\mathbf{y}$. $\mathcal{N}(\mathbf{x};\bm{\mu},\bm{\Sigma})$ means $\mathbf{x}$ following a Gaussian distribution with mean $\bm{\mu}$ and covariance matrix $\bm{\Sigma}$. ${\left\| {\bf{A}} \right\|_F}$ and $[{\bf{A}}]_{m,n}$ denote the Frobenius norm and the $m$-th row and $n$-th column element of the matrix ${\bf{A}}$, respectively. $\| {\bf x} \|_p$, $[{\bf x}]_{m}$, and $|{\bf x}|_c$ denote the ${l_p}$ norm, the $m$-th element, and the cardinality of the vector ${\bf x}$, respectively. $\mathbf{I}$ is the identity matrix. $\bf{0}$ is a vector with all the elements being 0.

\vspace{-2mm}
\section{Proposed Swin Transformer-Based Downlink CSI Acquisition Scheme}\label{S2}

\subsection{System Model}\label{S2.1}
\vspace{-1mm}
We assume that the BS deploys a uniform planar array (UPA) with $N_{\rm{BS}}$ antennas to serve $U$ single-antenna UEs in an FDD mode. Orthogonal frequency division multiplexing (OFDM) with $P$ subcarriers is considered. To realize the downlink CSI acquisition, the BS first broadcasts the downlink pilot signal to the UEs, and then each UE estimates the CSI and feeds it back to the BS. We focus on the CE and feedback of a single UE, and its received signal ${\bf{y}}_p\! \in\! \mathbb{C}^{ M}$ on the $p$-th subcarrier in $M$ successive time slots can be expressed as 
	\vspace{-0.5mm}
\begin{equation}\label{fd:received-signal} 
	{\bf{y}}_p^{\rm T} = {\bf{h}}_p^{\rm T} {\bf{X}} + {\bf{n}}_p^{\rm T},
	\vspace{-0.5mm}
\end{equation}
where ${\bf{X}}\! \in\! \mathbb{C}^{N_{\rm{BS}} \times M}$ is the transmit signal in $M$ successive time slots, ${\bf{h}}_p\! \in\! \mathbb{C}^{N_{\rm{BS}}}$ is the $p$-th subcarrier's channel, and ${\bf{n}}_p^{\rm T}$ is complex additive white Gaussian noise (AWGN) with zero mean and covariance matrix $\sigma_n^2{\bf{I}}$. By stacking the received signals over all the subcarriers, the received signal ${\bf{Y}}\! =\! \left[ {\bf{y}}_1, {\bf{y}}_2, \cdots, {\bf{y}}_P \right]^{\rm T}\! \in\! \mathbb{C}^{P \times M}$ can be written as
\vspace{-0.5mm}
\begin{equation}\label{frequency-spatial} 
	{\bf{Y}} = {\bf{H}}_s {\bf{X}} + {\bf{N}},
		\vspace{-0.5mm}
\end{equation}
where ${\bf{H}}_s = \left[ {\bf{h}}_1, {\bf{h}}_2, \cdots, {\bf{h}}_P \right]^{\rm T}\! \in\! \mathbb{C}^{P \times N_{\rm{BS}}}$ is the frequency-spatial domain channel matrix, and ${\bf{N}} = \left[ {\bf{n}}_1, {\bf{n}}_2, \cdots, {\bf{n}}_P \right]^{\rm T}$ is the AWGN matrix. Furthermore, we can obtain the frequency-angle domain channel ${\bf{H}}_a\! \in\! \mathbb{C}^{P \times N_{\rm{BS}}}$ as \cite{gao2015spatially}
\vspace{-0.5mm}
\begin{equation}\label{angulr} 
	{\bf{H}}_a = {\bf{H}}_s {\bf{F}},
	\vspace{-0.5mm}
\end{equation}
where ${\bf{F}}\! \in\! \mathbb{C}^{N_{\mathrm{BS}} \times N_{\rm{BS}}}$ a discrete Fourier transform (DFT) matrix. Due to the angular-domain sparsity, each row of ${\bf{H}}_a$ is a sparse vector. Hence, (\ref{frequency-spatial}) can be rewritten as
\vspace{-0.5mm}
\begin{equation}\label{problem} 
	{\bf{Y}} = {\bf{H}}_{a} {\bf{F}}^{\rm H} {\bf{X}} + {\bf{N}} = {\bf{H}}_{a} {\widetilde{\bf{X}}} + {\bf{N}},
	\vspace{-0.5mm}
\end{equation}
where ${\widetilde{\bf{X}}} = {\bf{F}}^{\rm H} {\bf{X}}$.

\vspace{-3mm}
\subsection{Pilot Design and CSI Acquisition}\label{S2.2}

The structure of the proposed SWTCAN is shown in Fig.~\ref{fig1}. We use a linear layer without bias to model the pilot design. The received pilot signal is passed through a linear layer, and its dimension is restored to be the same as the channel.  

The core of the compressor consists of 8 Swin Transformer blocks\cite{swin}. These blocks are responsible for extracting high-dimensional features from the input signal, leveraging a 96-dimensional embedding space. Each pair of Swin Transformer blocks starts with LayerNorm, followed by shifted window multi-head self-attention (SW-MSA) in the first block and  window-based multi-Head self-attention (W-MSA) in the second, both paired with multilayer perceptron (MLPs) and residual connections. This structure captures the local and global features of CSI effectively. The features passes through a linear layer with an activation function and form a codeword.
\color{black}
The compressed codeword is then quantized into $B$ bits vector $\mathbf{q}$ by a quantization layer. The entire CSI compressor can be expressed as
\begin{equation}\label{equ5} 
	\mathbf{q} = \mathcal{Q} \left( f_{\downarrow}\left( \mathbf{Y},\bm{\theta}_{\downarrow}\right) \right) \in \mathbb{R}^{B },
\end{equation} 
where $\mathcal{Q}(\cdot)$, $f_{\downarrow}(\cdot ,\cdot )$, and $\bm{\theta}_{\downarrow}$ denote the quantization function, compression function and learnable neural network parameters in the compressor, respectively.
		
The structure of the CSI reconstructor at the BS is similar to that of the compressor. The received bit vector is transformed by a dequantization layer and a linear layer to change the feature dimension, which is then inputted to Swin Transformer blocks. Finally, the features are up-sampled by a patch expanding layer to reconstruct the downlink CSI $\widehat{\mathbf{H}}_{a}$ at the BS. The entire CSI reconstructor can be expressed as
\begin{equation}\label{eqRec} 
	\widehat{\mathbf{H}}_{a} = f_{\uparrow}\left( \mathcal{Q}^{-1} \left( \mathbf{q}\right),\bm{\theta}_{\uparrow}\right), 
\end{equation}
where $f_{\uparrow}(.,.)$, $\mathcal{Q}^{-1}(\cdot)$, and $\bm{\theta}_{\uparrow}$ denote the reconstruction function, dequantization function and learnable neural network parameters, respectively. By adopting the normalized mean squared error (NMSE) $L_{1}\! =\! \frac{\left\| \widehat{\mathbf{H}}_{a}-\mathbf{H}_{a}\right\|_{F}^2}{\left\| \mathbf{H}_{a}\right\|_{F}^2}$  as the loss function, we can perform E2E training on the proposed SWTCAN.
	
\vspace{-1.5mm}	
\section{Generative Pre-Training and Federated-Tuning for the Proposed SWTCAN}\label{section3}
 
Our proposed CSI-GPT framework integrates the proposed VAE-CSG to pre-train SWTCAN with a small number of CSI samples from the current cell. We also adopt a FL-based online fine-tuning to further improve the performance of pre-trained SWTCAN, which has much lower communication overhead than the CL scheme. The procedure of CSI-GPT with federated-tuning is summarized in Algorithm~\ref{algorithm1}.

\vspace{-4mm}
\subsection{Generative AI-Based Pre-Training}\label{S3.1}

The proposed VAE-based generative network for generating CSI samples, called VAE-CSG, pre-trains the SWTCAN so that it can initially learn the generalized CSI features before fine-tuning it, which helps the model to perform better in the subsequent task and accelerate the convergence speed. As shown in line~\ref{line1} of Algorithm~\ref{algorithm1}, the BS initially pre-trains the VAE-CSG using a large amount of CSI samples generated by the channel simulator, and these samples typically have a different channel distribution from the current cell. Then the VAE-CSG is fine-tuned using a limited number of CSI samples from the current cell, which are obtained from the uplink CE at a high signal-to-noise ratio (SNR)\footnote{Although the reciprocity of uplink and downlink channels does not hold in FDD, they usually have similar distributions and features. However, due to the limited transmit power of UEs, the uplink SNR is usually low, and high-quality uplink CSI samples at high SNR are limited.}.

\begin{algorithm}[t] 
\small
\caption{Proposed CSI-GPT with Federated-Tuning}
\label{algorithm1}
\begin{algorithmic}[1]
	\STATE BS pre-trains VAE-CSG with simulated CSI samples, and fine-tunes VAE-CSG  with limited CSI samples obtained in current cell to generate more CSI samples. \label{line1}
	\STATE BS pre-trains SWTCAN with data generated by VAE-CSG. \label{line2}
	\STATE Initialize federated-tuning parameters $\mathbf{m}_0 = 0$, $\mathbf{v}_0 = 0$. \label{line3}
	\FOR {$t = 1,2, \cdots, T$} \label{line4}
		\STATE BS broadcasts $\tilde{\bm{\theta}}_{t}$ to all UEs.\label{line5}
			\FOR {each UE $u \in \mathcal{S}$ \textbf{in parallel}} \label{line6}
				\STATE Initialize local parameters $\tilde{\bm{\theta}}^{u,0}_{t} =\tilde{\bm{\theta}}_{t}$. \label{line7}
				\STATE Use SGD for local model updates, according to (\ref{equ9}).\label{line8}
		\ENDFOR \label{line9}
		\STATE Perform AirComp, according to (\ref{equ10}).\label{line10}
		\STATE BS updates the  model parameter $\tilde{\bm{\theta}}_{t+1}$, according to (\ref{equ11}).\label{line11}
	\ENDFOR\label{line15}
\end{algorithmic}
\end{algorithm}
\setlength{\textfloatsep}{2pt}

The VAE-CSG comprises an encoder and a decoder, which are jointly trained using an E2E method and only the decoder is utilized for generating CSI samples. Denote the parameters of the encoder and decoder of VAE-CSG as ${\bm{\psi}}$ and ${\bm{\omega}}$, respectively. The output of the encoder, i.e., the latent variable, is denoted as $\mathbf{z}\! =\! f_{enc}(\mathbf{H}_{a};\bm{\psi})$. Similarly, the output of the decoder is denoted as $f_{dec}(\mathbf{z};\bm{\omega})$. The loss function of the VAE-CSG consists of two parts. The first part is the reconstruction loss, $\|f_{dec}(\mathbf{z};\bm{\omega})\! -\! \mathbf{H}_{a}\|_F^2$, which measures the closeness of the VAE-CSG's output to the original input. According to \cite{kingma2013auto}, the second part measures the difference between the learned distribution of latent variable $\mathbf{z}$ and the predefined prior distribution $p_0(\mathbf{z})\! \sim\! \mathcal{N}(\mathbf{z};\mathbf{0},\mathbf{I})$. Let the learned distribution of $\mathbf{z}$ be $p_{\bm{\psi}}(\mathbf{z}|\mathbf{H}_{a})$. Using Kullback-Leibler (KL) divergence, the second part of the loss is ${\text D}_{KL}(p_{\bm{\psi}}(\mathbf{z}|\mathbf{H}_{a})||p_0(\mathbf{z}))$. Hence, the loss function of VAE-CSG is expressed as
\begin{align}\label{loss2} 
	L_{2}(\bm{\psi},\bm{\omega}) =& \|f_{dec}(\mathbf{z},\bm{\omega})-\mathbf{H}_{a}\|_F^2 \nonumber \\
	& + l \cdot {\text D}_{KL}(p_{\bm{\psi}}(\mathbf{z}|\mathbf{H}_{a})||p_0(\mathbf{z})) ,
\end{align}
where \textit{l} is a predefined hyper-parameter. The second term in (\ref{loss2}) enhances the diversity and quality of the generated CSI samples by enforcing the encoder to produce a latent variable that closely matches the standard Gaussian distribution. We use the data generated by VAE-CSG to pre-train SWTCAN (line \ref{line2} in Algorithm~\ref{algorithm1}). The performance of pre-trained SWTCAN is suboptimal, since the CSI distributions used in pre-training and testing in practical deployment are usually different.	
		
\vspace{-4mm}
\subsection{Federated Learning-Based Online Fine-Tuning}\label{S3.2}
		
To enhance the performance of the pre-trained SWTCAN, fine-tuning is necessary. Adopting the CL strategy for fine-tuning would require the BS to aggregate a large number of CSI samples from UEs in the current cell, resulting in excessive uplink communication overhead. The downlink CSI used for fine-tuning can be obtained by the BS broadcasting the pilot signals, facilitating each UE to estimate its own downlink CSI based on the pilot signals. To address the high communication overhead and privacy issue  caused by feeding these CSI samples back to the BS, we utilize the communication-efficient federated-tuning and AirComp to fine-tune the SWTCAN, while reducing uplink communication costs and overhead.
	
\subsubsection{Communication-Efficient Federated-Tuning}
 
Uploading the entire parameter set $\bm{\theta}$ of SWTCAN for fine-tuning would result in prohibitive  uplink communication overhead. We opt to freeze the majority parameters of SWTCAN and upload only a minority of the entire parameters, denoted by $\tilde{\bm{\theta}}$. Hence, in federated-tuning, we solve the optimization:
\begin{equation}\label{eqFLopt} 
	\min\nolimits_{\tilde{\bm{\theta}} \in \mathbb{R}^{d}} L_{1}(\tilde{\bm{\theta}})=\frac{1}{U} \sum\nolimits_{u=1}\nolimits^{U} L_{1}^u(\tilde{\bm{\theta}}),
\end{equation}
where $d\! =\! |\tilde{\bm{\theta}}|_c $ is the dimension of the learnable parameters and $L_{1}^u(\tilde{\bm{\theta}})$ is the NMSE loss function of the $u$-th UE.

\subsubsection{AirComp for Efficient Federated-Tuning}
To accelerate federated-tuning convergence and minimize the uplink communication rounds, we employ the federated AMSGrad with max stabilization (FedAMS) \cite{fedams} in conjunction with AirComp. 

The BS begins by initializing the SWTCAN model with pre-trained parameters. These parameters have been pre-trained using a large dataset of CSI samples generated by the VAE-CSG.
\color{black}
In the $t$-th communication round, $1\le t\le T$, the BS broadcasts the learnable model parameter $\tilde{\bm{\theta}}_{t}$ to all UEs (line \ref{line5} in Algorithm~\ref{algorithm1}). Due to the heterogeneous user availability, only a fraction of UEs, denoted as $\mathcal{S}$, participates in the $t$-th communication round.	The $u$-th UE, $\forall u\! \in\! \mathcal{S}$, initializes its parameters as $\tilde{\bm{\theta}}^{u,0}_{t}\! =\!\tilde{\bm{\theta}}_{t}$, and then minimizes its local loss function by conducting $K$ local training epochs with the local learning rate $\eta_l$ through its local dataset (lines \ref{line6}--\ref{line9}). In the $k$-th local training epoch, $1\le k\le K$, learnable parameters $\tilde{\bm{\theta}}^{u,k}_{t}$ can be updated by stochastic gradient descent (SGD), which is expressed as
\begin{equation} \label{equ9} 
	\tilde{\bm{\theta}}^{u,k}_{t}=\tilde{\bm{\theta}}^{u,k-1}_{t}-\eta_{l} \nabla L_{1}^u(\tilde{\bm{\theta}}^{u,k-1}_{t}),
	\vspace{-1mm}
\end{equation}
where $\nabla$ denotes the gradient operator. 
After $K$ local training epochs, instead of sending the entire model back to the BS, the $u$-th UE sends the model difference $\Delta \tilde{\bm{\theta}}^{u}_{t} = \tilde{\bm{\theta}}^{u,K}_{t}\! -\! \tilde{\bm{\theta}}^{u,0}_{t}$ to the BS, and the BS receives the sum of the local model differences from multiple devices based on AirComp (line \ref{line10} in Algorithm~\ref{algorithm1}), which can be expressed as 
\begin{equation}\label{equ10} 
	\bm{\delta}_{t} = \frac{1}{|\mathcal{S}|_c} \sum\nolimits_{u\in \mathcal{S}} \Delta \tilde{\bm{\theta}}^{u}_{t}  + \mathbf{n}^{+},	
	\vspace{-1mm}	
\end{equation}
where $\mathbf{n}^{+}\! \in\! \mathbb{R}^{d}$ is the noise in the uplink aggregation process, $\bm{\delta}_{t}$ represents the noisy model difference, which is also treated as a pseudo gradient to update the model at the BS.
\color{black} According to \cite{fedams}, the BS updates the learnable parameters $\tilde{\bm{\theta}}_{t+1}$ for the ($t$+1)-th round (line \ref{line11} in Algorithm~\ref{algorithm1}) as
\begin{equation}\label{equ11} 
	\tilde{\bm{\theta}}_{t+1} = \tilde{\bm{\theta}}_{t}+\eta \frac{\mathbf{m}_{t}}{\sqrt{\mathbf{v}_{t}}},	
	\vspace{-1mm}
\end{equation}
where $\eta$ is the global learning rate, $\mathbf{m}_{t}$ is the momentum and $\mathbf{v}_{t}$ is the variance in the $t$-th round, which are updated by $\mathbf{m}_{t}\! =\! \beta_{1} \mathbf{m}_{t-1}\! +\! (1\! -\! \beta_{1}) \bm{\delta}_{t}$ and $\mathbf{v}_{t}\! =\! \max \left\{ \mathbf{v}_{t-1}\! +\! (1\! -\!\beta_{2}) \bm{\delta}_{t}^{2}, \mathbf{v}_{t-1}\right\}$ with the hyperparameters $\beta_1$ and $\beta_2$.

\vspace{-2mm}
\section{Simulation Results}\label{S4}
		
We use the Sionna library in Python to generate MIMO channel realizations. The training and test datasets contain the same number of various types of CDL channels, namely, CDL-A, CDL-B, and CDL-C, where the CDL channel models are adopted from 3GPP standards \cite{3gpptr38901}. The sizes of the training, validation, and test channel datasets  are 6000, 2000, and 2000, respectively. The carrier frequency is 28\,GHz, the number of subcarriers is $P\! =\! 256$ and the subcarrier spacing is 240\,kHz. The BS is equipped with the UPA of $N_{\mathrm{BS}}\! =\!16\times 16\! =\!256$ antennas, with antenna space $\frac{\lambda_{c}}{2}$, where $\lambda_{c}$ is the signal wavelength. The delay spread is 30\,ns, and the downlink SNR is 20\,dB.
\begin{figure}[t]
	\vspace{-6mm}
	\begin{center}
		\subfloat[$\rho=8$] {\includegraphics[width=0.7\columnwidth]{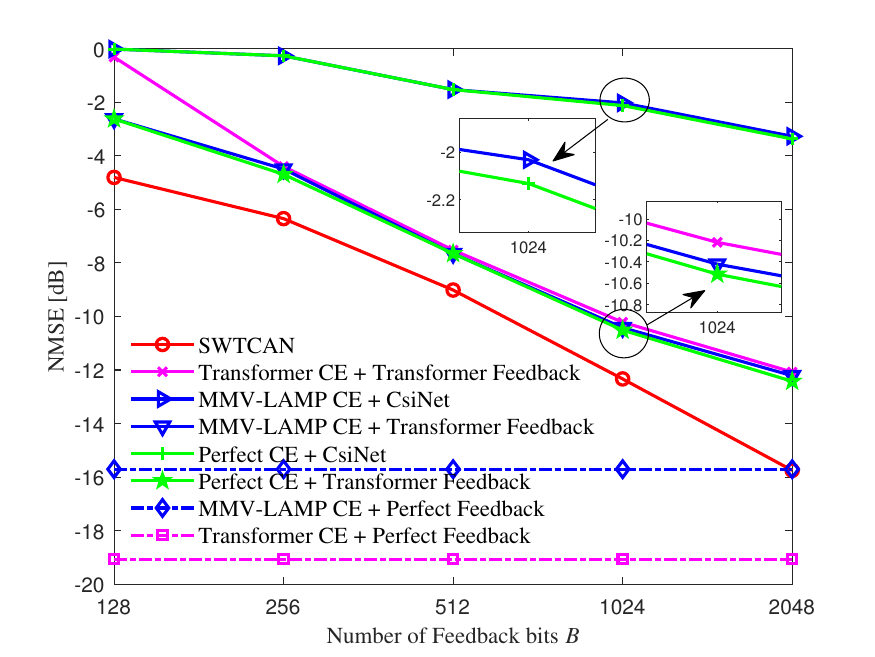}\label{fig4a}}
		\\
		\vspace{-4mm}
		\subfloat[$\rho=16$]{\includegraphics[width=0.7\columnwidth]{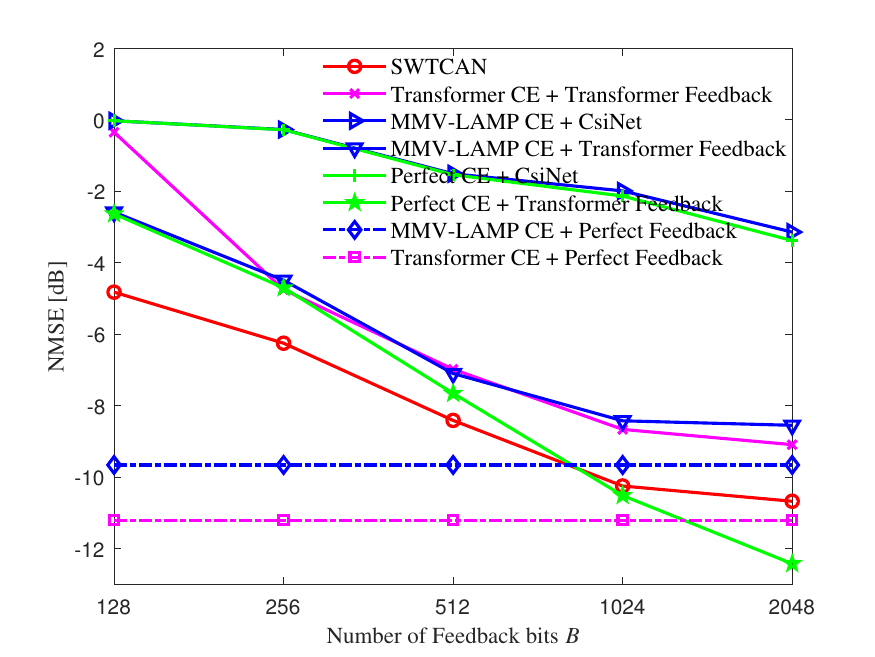}\label{fig4b}}
	\end{center}	
	\vspace{-4mm}
	\caption{NMSE performance of different schemes versus the feedback overhead $B$ for CDL-B.} 
	\label{fig5} 
\end{figure}

\begin{figure*}[h]
	\vspace{-0mm}
	\centering
	\subfloat[$\rho=2$]{\includegraphics[width=0.35\textwidth]{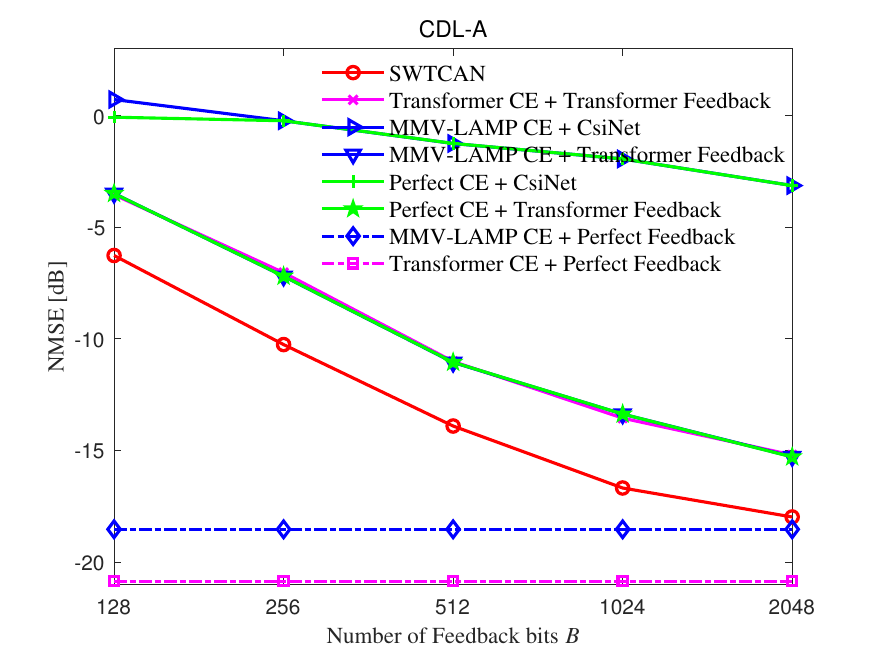} \label{fig5a}}
	\subfloat[$\rho=2$]{\includegraphics[width=0.35\textwidth]{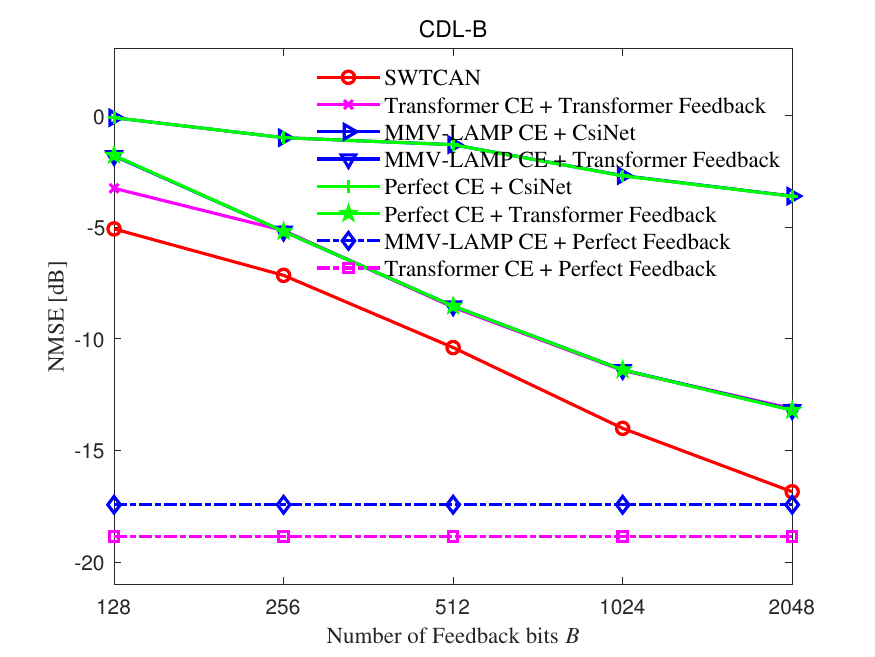}\label{fig6a}}
	\subfloat[$\rho=2$]{\includegraphics[width=0.35\textwidth]{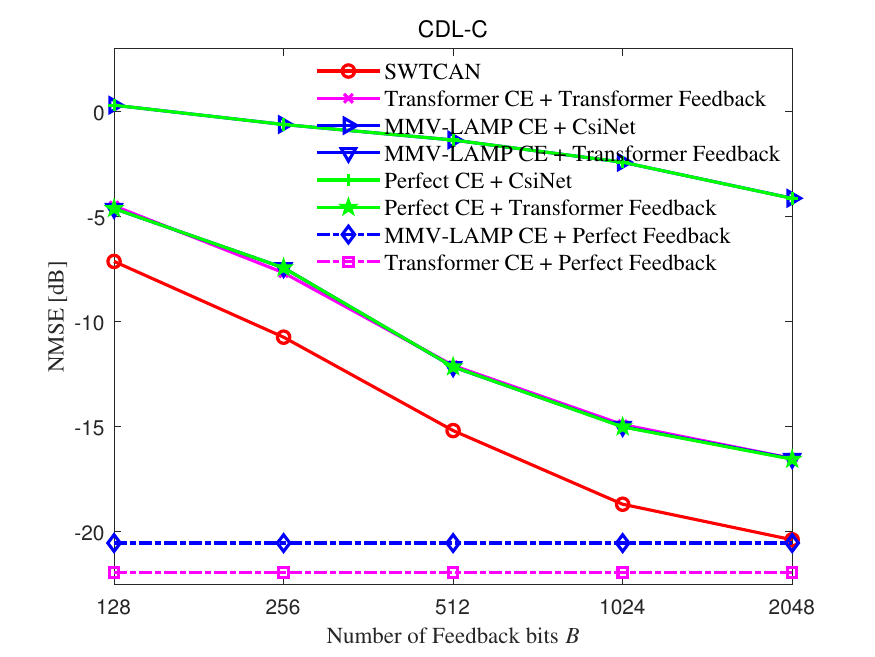}\label{fig7a}}
	
	\quad
	\subfloat[$\rho=32$]{\includegraphics[width=0.35\textwidth]{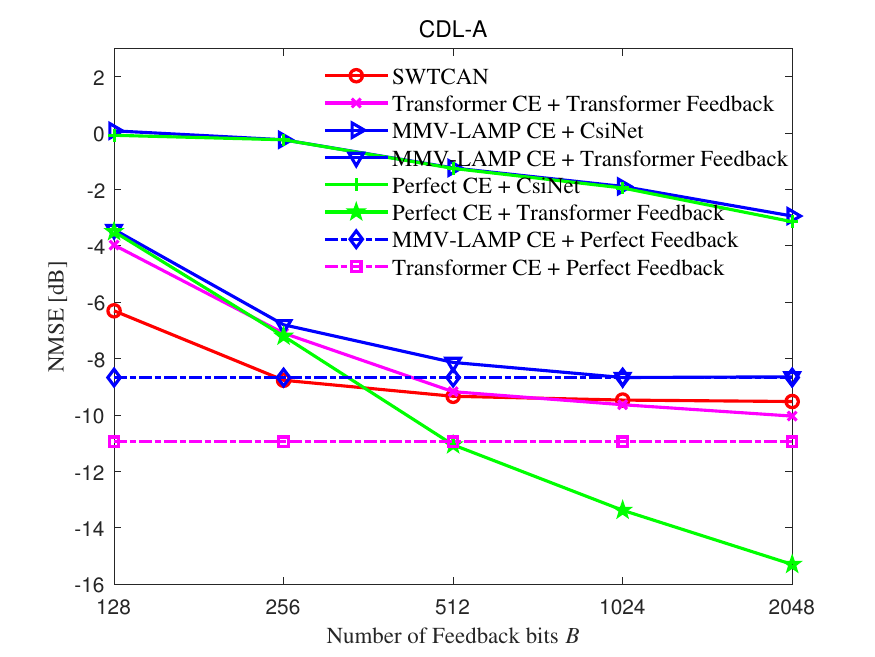}\label{fig5b}}
	\subfloat[$\rho=32$]{\includegraphics[width=0.35\textwidth]{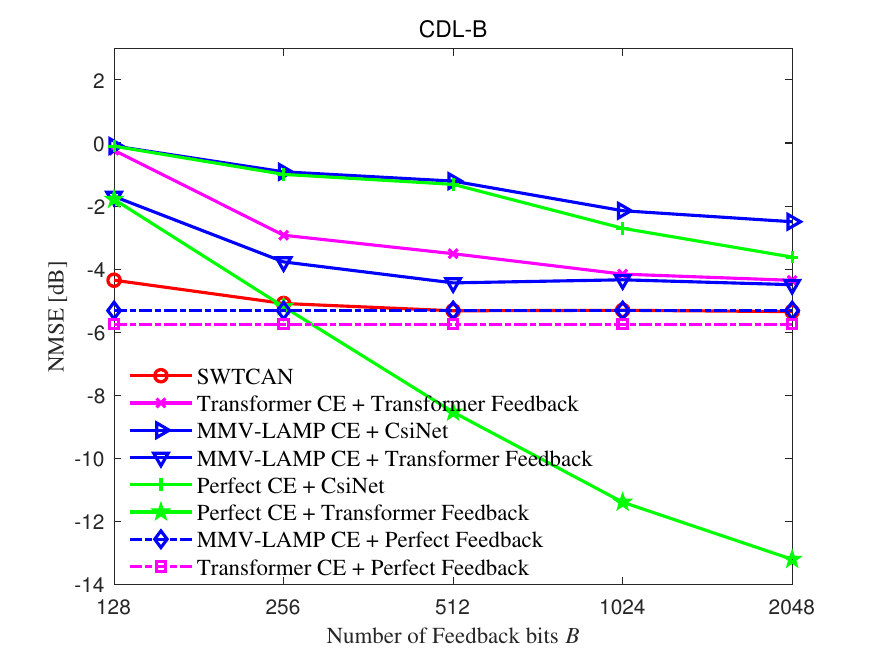}\label{fig6b}}
	\subfloat[$\rho=32$]{\includegraphics[width=0.35\textwidth]{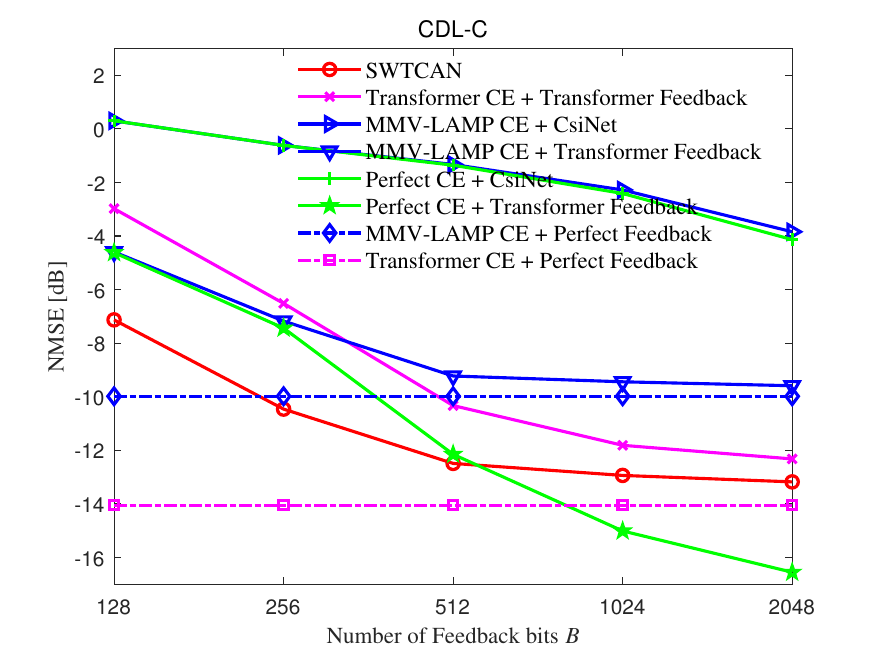}\label{fig7b}}
	\captionsetup{justification = raggedright,labelsep=period}
	\caption{NMSE performance of different schemes versus the feedback signaling overhead for CDL-A/B/C. } \label{fig5-1}
	\vspace{-5mm}
\end{figure*}

\vspace{-3mm}
\subsection{Performance of Proposed SWTCAN}\label{S4.1}
		
We compare the proposed SWTCAN with the following baselines.
\textbf{Baseline~1}: The Transformer-based network for CE and CSI feedback \cite{wangyang_magzine}, denoted as 		`Transformer CE\,+\,Transformer Feedback'. 
\textbf{Baseline~2}: The multiple-measurement-vectors (MMV)-learned approximate message passing (LAMP) algorithm \cite{xisuo_estiamtion_feedback} for CE and the bit-level CsiNet scheme with an attention mechanism \cite{bitcsinet} or a Transformer-based network for CSI feedback, denoted as `MMV-LAMP CE\,+\,CsiNet/Transformer Feedback'.
\textbf{Baseline~3}: The perfect downlink channel estimate and the bit-level CsiNet scheme with an attention mechanism \cite{bitcsinet} or a Transformer-based network for CSI feedback, denoted as `Perfect CE\,+\,CsiNet/Transformer Feedback'.
\textbf {Baseline~4}: The MMV-LAMP algorithm or a Transformer-based network for CE and the perfect CSI feedback to the BS, denoted as `MMV-LAMP/Transformer CE\,+\,Perfect Feedback'. 

Fig.~\ref{fig5} shows the NMSE performance achieved by different schemes versus the feedback overhead \textit{B}. Both the training and test datasets are CDL-B channels. 
Fig.~\ref{fig4a} indicates that at a compression ratio $\rho\! =\! \frac{N_{\rm{BS}}}{M}\! =\! 8$, the main factor influencing the performance is the CSI feedback scheme. Our SWTCAN demonstrates excellent performance across all feedback bit numbers \textit{B}, outperforming the baseline schemes and approaching the performance of perfect CSI feedback at $B\! =\! 2048$. 
Fig.~\ref{fig4b} shows that at a higher $\rho\! =\! 16$, the CE algorithm also affects the final performance. 
Simulations on CDL-A and CDL-C channels, not showing due to space limits, also verify the same advantages of our proposed SWTCAN.
Thank you for your suggestion regarding the inclusion of additional experiments with different channel configurations. We agree that demonstrating the generalizability of our proposed approach across various channel models is important. While our primary simulations were conducted under the CDL-B channel, we have also performed extensive simulations using CDL-A and CDL-C channels. These additional experiments confirm the superiority of our proposed algorithm across different channel conditions. However, due to the page limit constraints (with a maximum of 7 figures and tables allowed), we were unable to include these results in the main manuscript. We have attached the relevant simulation results and analysis in this response letter for your review.

As shown in Fig.~\ref{fig5-1}, we extended our simulations to include compression ratios $\rho$ of 2 and 32, in addition to the $\rho$ = 8 and $\rho$ = 16 scenarios presented in the manuscript. These extended simulations were conducted across CDL-A, CDL-B, and CDL-C channels to examine the boundary effects on overall performance, specifically regarding channel estimation and feedback algorithms. 
The results showed that at a compression ratio of $\rho$ = 2, the performance is primarily determined by the channel feedback algorithm. Our proposed SWTCAN algorithm approaches the performance of perfect CSI feedback when using the MMV-LAMP channel estimation algorithm. Conversely, at a compression ratio of $\rho$ = 32, channel estimation becomes the decisive factor for overall performance. Here, the SWTCAN algorithm outperforms the baseline in nearly all conditions, even with lossy channel estimation and feedback.
Notably, in the CDL-B channel at $\rho$ = 32, all algorithms requiring channel estimation demonstrate weaker performance, making the Transformer-based feedback scheme with perfect channel estimation particularly effective. In contrast, the CsiNet network, due to its less robust feature extraction capability, fails to achieve satisfactory performance even under perfect channel estimation.

In addition, we have analyzed the complexity of the proposed algorithm and the baselines. The complexity of the proposed SWTCAN mainly comes from (S)W-MSA layers, i.e.,  $\mathcal{O}\left( L P  N_{\mathrm{BS}} C^2 \!+\! L W^2 P  N_{\mathrm{BS}}C  \right) \approx \num{1.4e7},\!$ where $L$ is the number of layer in NN, $\!C\!$ is the number of channels of CSI, and $W$ is the size of windows. 
The complexity of  Transformer mainly comes from self-attention layers, i.e., $\mathcal{O}\left(L P^2 d_{model} \right) \approx \num{1.5e8} $, where $d_{model}$ is the dimension of linear embedding in Transformer.  The complexity of the MMV-LAMP algorithm mainly comes from matrix multiplication operations, i.e., $\mathcal{O}\left(G  M  P  N_{\mathrm{BS}}^2 I\right)\approx \num{1.1e10} $, where $G$ is the oversampling factor in redundant dictionary,  $I$ is the number of iterations.  The complexity of CsiNet mainly comes from convolutional layers, i.e., $\mathcal{O}\left(  P  N_{\mathrm{BS}} N_{co}^2\sum_{i=1}^{L} n_{i-1} n_{i} \right)\approx \num{2.4e6} $, where $N_{co}$  is the size
of the convolutional filters, $n_{i-1}$ and $ n_{i}$ are the numbers of input and output feature maps of the $i$-th convolutional layer. 
The values of the above parameters can be obtained from the open source code.
It is evident that the proposed SWTCAN exhibits lower complexity than the baselines while maintaining good performance.
\color{black}
		
\vspace{-3mm}
\subsection{Performance of Proposed Generative Pre-Training}\label{S4.2}

We evaluate the performance of generative pre-training on SWTCAN at $\rho\! =\! 8$ and $B\! =\! 512$, 1024 and 2048 bits with the value of $l$ in (\ref{loss2}) set to 0.00025. We assume that the BS has 6000 CDL-A CSI samples, which are obtained from the channel model generator. But the true CSI distribution in the BS's cell follows a different CDL-B distribution. The BS only has 120 high-quality CDL-B samples, which are obtained from the uplink CE. We compare the proposed scheme and three benchmark schemes for ablation study. \textbf{Scheme~A}: We pre-train SWTCAN with 6000 CDL-A samples. \textbf{Scheme~B}: We pre-train SWTCAN with 120 CDL-B samples. 
\textbf{Scheme~C}: We pre-train SWTCAN with 6000 CDL-A samples, fine-tune it with 120 CDL-B samples.
\color{black}
\textbf{Scheme~D}: We train VAE-CSG with 120 CDL-B samples, and then use it to generate 6000 CSI samples to pre-train SWTCAN. By contrast, in the \textbf{Proposed} scheme, we pre-train VAE-CSG with 6000 CDL-A samples, fine-tune it with 120 CDL-B samples, and finally generate 6000 CSI samples to pre-train SWTCAN. These four pre-training schemes are tested on 2000 CDL-B samples, and the pre-training test NMSEs are compared in Table~\ref{part2}. It can be seen that the proposed VAE-CSG outperforms the other three benchmark schemes, demonstrating its effectiveness. Based on the proposed pre-training strategy, the performance of SWTCAN at $\rho\! =\! 8$ and $B\! =\! 2048$ bits reaches the NMSE of -9.3678\,dB, which is still a bit short of -15.7\,dB shown in Fig.~\ref{fig4a}. Therefore, we use federated-tuning to further improve the performance.
	
\begin{table}[t]
\caption{Pre-training test NMSE (dB) performance comparison of different pre-training schemes under the CDL-B channel.}
\label{part2} 
\vspace*{-3mm}
\begin{center}
\resizebox{1\columnwidth}{!}{
\begin{tabular}{|c|c|c|c|c|c|}
	\hline
	Number of feedback bits & {\bf Scheme A} & {\bf Scheme B} & {\bf Scheme C} \color{black} & {\bf Scheme D} & {\bf Proposed} \\ \hline
	\textit{B}=512          & 0.2009         & -2.8306        & -3.8199 \color{black}		& -4.0628        & \textbf{-5.0361} \\ \hline
	\textit{B}=1024         & 0.2491         & -3.7469        & -4.8259	\color{black}		& -5.5036        & \textbf{-7.3138} \\ \hline
	\textit{B}=2048         & -0.1258        & -3.9417        & -5.5449 	\color{black}	&-6.6860         & \textbf{-9.3678} \\ \hline
\end{tabular}}
\end{center}
\vspace*{-1mm}
\end{table}
		
\vspace{-2mm}
\subsection{Performance of Proposed Federated-Tuning Method}\label{S4.3}
		
In this ablation study, there are $U\! =\! 600$ UEs, each having $N_{\rm FL}^s=10$ actual CSI samples.	During each communication round of federated-tuning, 10\% of UEs, i.e., 60 UEs, are involved, and each UE conducts $K\! =\! 2$ local training epochs with the local training rate $\eta_{l}\! =\! 0.001$ to facilitate online fine-tuning of SWTCAN at $\rho\! =\! 8$ and $B\! =\! 2048$\,bits. The hyperparameters in (\ref{equ11}) are set to $\eta\! =\! 1$, $\beta_1\! =\! 0.9$, and $\beta_2\! =\! 0.99$. We freeze the parameters of the SWTCAN except for the last two layers
\footnote{
Our study showed that freezing the last two layers of all Swin Transformer Blocks (with 3,617,280 unfrozen parameters) resulted in a final performance of -10.722 dB after 25 global epochs. Freezing only the last layer (2,509,299 unfrozen parameters) led to a performance of -10.299 dB, while freezing half of the last layer (2,471,667 unfrozen parameters) achieved -10.202 dB. The chosen freezing scheme strikes a balance between communication efficiency and performance, which is verified in the simulations.
\color{black}} 
in the decoder. SNR is set to 20\,dB for both downlink CE and uplink AirComp. For federated-tuning, computation (updating the trainable parameters $\tilde{\bm{\theta}}$) is done in the UE, and the model updates of multiple UEs are transmitted using AirComp\footnote{Note that the uplink communication overhead can be further reduced using gradient compression methods \cite{amiri2020federated}. Due to space limitation, this point is not discussed in this paper, which will be investigated in future.}. For CL, the BS collects CSI samples via orthogonal transmission, and then updates the whole model parameters $\bm{\theta}$. Note that $|\tilde{\bm{\theta}}|_c \!= \!$ \numprint{3617280},  $|\bm{\theta}|_c \!= \!$ \numprint{32623524}, hence $|\tilde{\bm{\theta}}|_c \approx 0.11|\bm{\theta}|_c$.

\begin{table}[!h]
\vspace*{-1mm}
\caption{Characterization of communication overhead, computation cost and computation speed of federated-tuning and CL schemes.}
\label{table2} 
\vspace*{-3mm}
\begin{center}
\resizebox{1\columnwidth}{!}{
\begin{threeparttable}[b]
\begin{tabular}{|c|c|c|}
	\hline
	\diagbox{Items}{Schemes} & \begin{tabular}[c]{@{}c@{}}Federated-tuning\\ (one global epoch)\end{tabular} & \begin{tabular}[c]{@{}c@{}}CL\\ (one CSI sample)\end{tabular} \\ \hline
	\begin{tabular}[c]{@{}c@{}}Communication overhead\\ ($\times 10^5$  real numbers) \end{tabular} & $|\tilde{\bm{\theta}}|_c \approx 36$ & $2PN_{\mathrm{BS}}\approx1.3$ \\ \hline
	Computation cost ($\times 10^9 $ FLOPs) \footnotemark[3] & $\zeta_{\mathrm{UE}} \approx 2.6$ & $\zeta_{\mathrm{BS}} \approx 3.4$ \\ \hline
	Computational speed (FLOPs/s) & $\kappa_{\mathrm{UE}}$ & $\kappa_{\mathrm{BS}} = \gamma \kappa_{\mathrm{UE}}$  \\ \hline			
\end{tabular}
\begin{tablenotes}
\item[3] The computation cost containing both forward and backward propagations is numerically calculated using torch-summary\cite{torchsummary}. Note that federated-tuning and CL have the same forward propagation, while federated-tuning has less computation cost in the backward propagation.
\end{tablenotes}
\end{threeparttable}
}
\end{center}
\vspace*{-3mm}
\end{table}

Table~\ref{table2} characterizes the communication overhead, computation cost and computation speed of the federated-tuning (one global epoch) and the CL scheme (one CSI sample). For federated-tuning, communication overhead during each global epoch is denoted as $|\tilde{\bm{\theta}}|_c$, representing the number of trainable parameters of SWTCAN. While, the communication overhead of CL is calculated as $2PN_{\mathrm{BS}}$ for each CSI sample. The computation cost is measured in FLOPs using the torch-summary tool, where $\zeta_{\mathrm{UE}}$ and $\zeta_{\mathrm{BS}}$ represent the computation cost in federated-tuning and CL, respectively. Computation speed, in FLOPs/s, is indicated by $\kappa_{\mathrm{BS}}$ and $\kappa_{\mathrm{UE}}$ for the BS and UE, respectively, where their ratio $\gamma = \frac{\kappa_{\mathrm{BS}}}{\kappa_{\mathrm{UE}}}$ reflects the computational power difference between the BS and UEs. 

For a fair comparison, later we will compare the NMSE performance of the two schemes, given the same communication resources and the same computation time. Here, we first calculate how many CSI samples the BS can collect in CL, given a fixed communication resource. Specifically, consider $T_0$ global epochs of federated-tuning, within which period the communication overhead is $T_0|\tilde{\bm{\theta}}|_c$. Using the same communication resources, the BS in CL can collect $N_{\rm CL}^s\! =\! \frac{T_0|\tilde{\bm{\theta}}|_c}{2PN_{\mathrm{BS}}}$ CSI samples for central training. Furthermore, we calculate how many training epochs CL can have, given a fixed computation time. In particular, given the number of global epochs $T_0$, the computation cost $\zeta_{\mathrm{UE}}$, the number of local CSI samples $N_{\rm FL}^s$, the number of local training epochs $K$, and the computation speed $\kappa_{\mathrm{UE}}$, the computation time of federated-tuning can be calculated as $\tau\! =\! \frac{T_0 N_{\rm FL}^s K \zeta_{\mathrm{UE}}}{\kappa_{\mathrm{UE}}}$. Within the same computation time $\tau$ in CL, given the computation cost $\zeta_{\mathrm{BS}}$, the number of CSI samples collected in the BS $N_{\rm CL}^s$, and the computation speed in the BS $\kappa_{\mathrm{BS}}$, we can obtain the number of training epochs of CL as $K_{\rm CL}\! =\! \frac{\tau \kappa_{\mathrm{BS}}}{ N_{\rm CL}^s \zeta_{\mathrm{BS}}}$. Although the BS is typically computationally more powerful than a UE, i.e., $\gamma > 1$, our proposed federated-tuning method still shows advantages due to the collaboration of multiple UEs.
\color{black}
\begin{figure}[h]
\vspace{-4mm}
\begin{center}
\includegraphics[width=0.7\columnwidth]{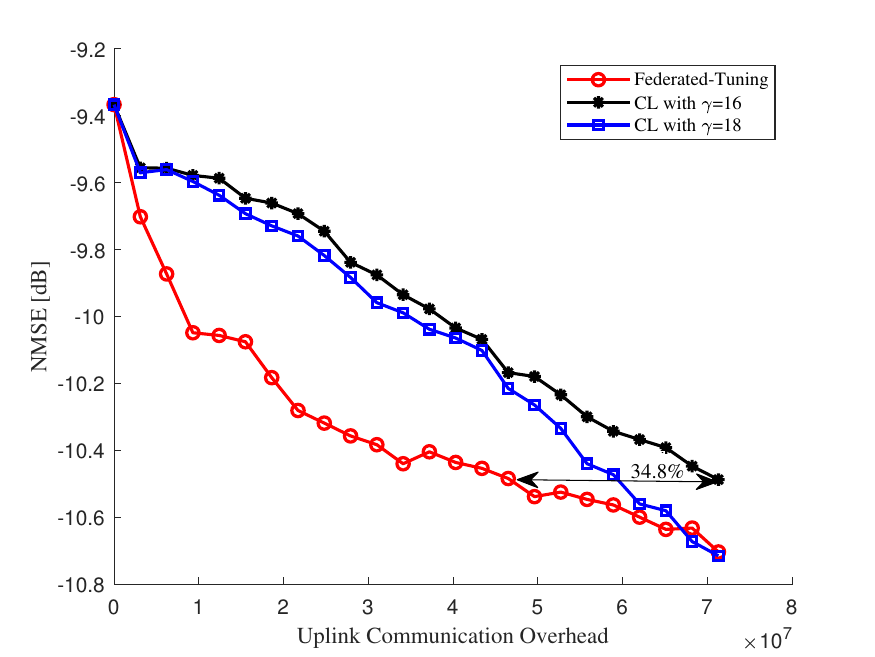}
\end{center}
\vspace{-4mm}
\caption{Performance comparison of federated-tuning and CL schemes versus the uplink communication overhead.}
\label{part3} 
\vspace{-3mm}			
\end{figure}
			
Fig.~\ref{part3} compares the NMSE of the proposed federated-tuning with that of the CL scheme versus uplink communication overhead. It can be seen that our federated-tuning method exhibits superior NMSE performance compared to the CL scheme under the same computation time and the same uplink communication overhead. Specifically, to achieve an equivalent NMSE performance in the same computation time, compared to the CL scheme with $\gamma\! =\! 16$ (this situation is possible, e.g., a Nvidia 3090Ti graphics card with a computing power of 41.6\,TFLOPs on the BS and an iPad Air with a computing power of 2.6\,TFLOPs on the M1 chip of the UE), our proposed algorithm reduces uplink communication overhead by up to 34.8\%. For the CL scheme, as the ratio of  computational speed $\gamma$ increases, the number of training epochs that BS can conduct also increase given the same computation time, resulting in improved performance. 

\vspace{-2mm}
\section{Conclusions}

We have proposed a Swin Transformer-based CSI acquisition network called SWTCAN to jointly design the pilot, CSI compression and CSI reconstruction. In order to solve the training data scarcity problem as the actual CSI samples are difficult to measure, we have designed the VAE-CSG to generate CSI samples for pre-training SWTCAN. The combination of VAE-CSG and SWTCAN constitutes the downlink CSI acquisition network based on a generative pre-trained Transformer at the BS. To further enhance the performance of the pre-trained SWTCAN, we have utilized the communication-efficient federated-tuning and AirComp to fine-tune the SWTCAN, while substantially reducing uplink communication overhead. Simulations have demonstrated that our proposed SWTCAN has better performance compared to the state-of-the-art schemes, and have verified the communication efficiency of the proposed federated-tuning method. 
 
However, DL methods still impose high complexity and memory requirements on UEs and the implementation of AirComp introduces practical issues that require future efforts.

\color{black}
		
\bibliographystyle{IEEEtran}
\vspace{-2mm}

\begin{thebibliography}{10}
\providecommand{\url}[1]{#1}
\csname url@samestyle\endcsname
\providecommand{\newblock}{\relax}
\providecommand{\bibinfo}[2]{#2}
\providecommand{\BIBentrySTDinterwordspacing}{\spaceskip=0pt\relax}
\providecommand{\BIBentryALTinterwordstretchfactor}{4}
\providecommand{\BIBentryALTinterwordspacing}{\spaceskip=\fontdimen2\font plus
\BIBentryALTinterwordstretchfactor\fontdimen3\font minus
  \fontdimen4\font\relax}
\providecommand{\BIBforeignlanguage}[2]{{%
\expandafter\ifx\csname l@#1\endcsname\relax
\typeout{** WARNING: IEEEtran.bst: No hyphenation pattern has been}%
\typeout{** loaded for the language `#1'. Using the pattern for}%
\typeout{** the default language instead.}%
\else
\language=\csname l@#1\endcsname
\fi
#2}}
\providecommand{\BIBdecl}{\relax}
\BIBdecl

\bibitem{liu2024} 
H.~Liu, \emph{et al.}, ``Near-space communications: The last piece of 6G space–air–ground–sea integrated network pzzle,'' \emph{ Space Sci Technol.}, 2024;4:0176.DOI:10.34133/space.0176.
\color{black}

\bibitem{gao2015spatially} 
Z.~Gao, L.~Dai, Z.~Wang, and S.~Chen, ``Spatially common sparsity based adaptive channel estimation and feedback for FDD massive MIMO,'' \emph{IEEE Trans. Signal Process.}, vol.~63, no.~23, pp.~6169--6183, 2015.

\bibitem{jiang2024} 
B.~Jiang, \emph{et al.}, ``Total and minimum energy efficiency tradeoff in robust multigroup multicast satellite communications,'' \emph{ Space Sci Technol.}, 2023;3:0059.DOI:10.34133/space.0059.

\bibitem{zhang2024} 
R. Zhang,  \emph{et al.}, ``Integrated Sensing and Communication With Massive MIMO: A Unified Tensor Approach for Channel and Target Parameter Estimation,'' \emph{IEEE Trans. Wireless Commun.}, vol.~23, no.~8, pp. 8571-8587, Aug. 2024, doi: 10.1109/TWC.2024.3351856. 

\bibitem{Xu2024} 
J.~Xu, L.~You, G.~C.~Alexandropoulos, X.~Yi, W.~Wang, and X.~Gao, ``Near-field wideband extremely large-scale {MIMO} transmissions with holographic metasurface-based antenna arrays,'' \emph{IEEE Trans. Wireless Commun.},  doi: 10.1109/TWC.2024.3387709. 

\bibitem{Zhu2023}
P.~Zhu, H.~Lin, J.~Li, D.~Wang, and X.~You, ``High-performance channel estimation for mmWave wideband systems with hybrid structures,'' \emph{IEEE Trans. Commun.}, vol.~71, no.~4, pp. 2503--2516, Apr. 2023.

\bibitem{Fang2024}
J.~Fang, P.~Zhu, J.~Li, F.-C.~Zheng, and X.~You, ``Cell-free mMIMO systems in short packet transmission regime: Pilot and power allocation,'' \emph{IEEE Trans. Veh. Techno.}, vol.~73, no.~6, pp. 8322--8337, June 2024.

\bibitem{Ye2024}
Y.~Ye, L.~You, J.~Wang, H.~Xu, K.-K.~Wong, and X.~Gao, ``Fluid antenna-assisted {MIMO} transmission exploiting statistical {CSI},'' \emph{IEEE Commun. Lett.}, vol.~28, no.~1, pp. 223--227, Jan. 2024.

\color{black}


\bibitem{mxs_tvt} 
X.~Ma and Z.~Gao, ``Data-driven deep learning to design pilot and channel estimator for massive MIMO,'' \emph{IEEE Trans. Veh. Techno.}, vol.~69, no.~5, pp.~5677--5682, 2020.

\bibitem{Mahdi_TWC_estimation} 
M.~B.~Mashhadi and D.~G{\"u}nd{\"u}z, ``Pruning the pilots: Deep learning-based pilot design and channel estimation for MIMO-OFDM systems,'' \emph{IEEE Trans. Wireless Commun.}, vol.~20, no.~10, pp.~6315--6328, 2021.

\bibitem{csinet} 
C.-K.~Wen, W.-T.~Shih, and S.~Jin, ``Deep learning for massive MIMO CSI feedback,'' \emph{IEEE Wireless Commun. Lett.}, vol.~7, no.~5, pp.~748--751, 2018.

\bibitem{csinet_plus} 
J.~Guo, C.-K.~Wen, S.~Jin, and G.~Y.~Li, ``Convolutional neural network-based multiple-rate compressive sensing for massive MIMO CSI feedback: Design, simulation, and analysis,'' \emph{IEEE Trans. Wireless Commun.}, vol.~19,
 no.~4, pp.~2827--2840, 2020.


\bibitem{Nerini2023}
M.~Nerini, \emph{et al.}, ``Machine learning-based {CSI} feedback with variable length in {FDD} massive {MIMO},'' \emph{IEEE Trans. Wireless Commun.}, vol.~22, no.~5, pp. 2886--2900, May 2023.

\bibitem{guo2022}
Y.~Cui, \emph{et al.}, ``Lightweight neural network with knowledge distillation for {CSI} feedback,'' \emph{IEEE Trans. Commun.}, vol.~72, no.~8, pp. 4917-4929, Aug. 2024.


\color{black}

\bibitem{guo2022overview} 
J.~Guo, C.-K.~Wen, S.~Jin, and G.~Y.~Li, ``Overview of deep learning-based CSI feedback in massive MIMO systems,'' \emph{IEEE Trans. Commun.}, vol.~70, no.~12, pp.~8017--8045, 2022.

\bibitem{Jiajia_estimation_feedback} 
J.~Guo, \emph{et al.}, ``Deep learning for joint channel estimation and feedback in massive MIMO systems,'' \emph{Digital Commun. Netw.}, vol.~10, no.~1, pp.~83--93, 2024.

\bibitem{guo2021canet} 
J.~Guo, C.-K.~Wen, and S.~Jin, ``CAnet: Uplink-aided downlink channel acquisition in FDD massive MIMO using deep learning,'' \emph{IEEE Trans. Commun.}, vol.~70, no.~1, pp.~199--214, 2022.





\bibitem{wangyang_magzine} 
Y.~Wang, \emph{et al.}, ``Transformer-empowered 6G intelligent networks: From massive MIMO processing to semantic communication,'' \emph{IEEE Wireless Commun.}, vol.~30, no.~6, pp.~127--135, 2023.

\bibitem{Transformer_survey} 
K.~Han, \emph{et al.}, ``A survey on vision transformer,'' \emph{IEEE Trans. Pattern Anal. Mach. Intell.}, vol.~45, no.~1, pp.~87--110, 2023.

\bibitem{3gpptr38901} 
3GPP, ``Study on channel model for frequencies from 0.5 to 100 {GHz},'' 3rd Generation Partnership Project (3GPP), Tech. Rep. TR 38.901, May 2017.

\bibitem{channelGan} 
H.~Xiao, W.~Tian, W.~Liu, and J.~Shen, ``ChannelGAN: Deep learning-based channel modeling and generating,'' \emph{IEEE Wireless Commun. Lett.}, vol.~11, no.~3, pp.~650--654, 2022.

\bibitem{yang2022federated} 
Z.~Yang, \emph{et al.}, ``Federated learning for 6G: Applications, challenges, and opportunities,'' \emph{Engineering}, vol.~8, pp.~33--41, 2022.

\bibitem{amiri2020federated} 
M.~M.~Amiri and D.~G{\"u}nd{\"u}z, ``Federated learning over wireless fading channels,'' \emph{IEEE Trans, Wireless Commun.}, vol.~19, no.~5, pp.~3546--3557, 2020.


\bibitem{qiao2024}
L. Qiao, Z. Gao, M. B. Mashhadi and D. Güundüz, ``Massive Digital Over-the-Air Computation for Communication-Efficient Federated Edge Learning,'' \emph{IEEE J. Sel. Areas Commun.}, doi: 10.1109/JSAC.2024.3431572. 
\color{black}

\bibitem{chen2022fedtune} 
J.~Chen, \emph{et al.}, ``FedTune: A deep dive into efficient federated fine-tuning with pre-trained Transformers,'' \emph{arXiv preprint arXiv:2211.08025}, 2022.

\bibitem{Ahmet_fl_estimation_risMIMO} 
A.~M.~Elbir and S.~Coleri, ``Federated learning for channel estimation in conventional and RIS-assisted massive MIMO,'' \emph{IEEE Trans. Wireless Commun.}, vol.~21, no.~6, pp.~4255--4268, 2022.

\bibitem{DLL_FL_estimation_ris} 
L.~Dai and X.~Wei, ``Distributed machine learning based downlink channel estimation for RIS assisted wireless communications,'' \emph{IEEE Trans. Commun.}, vol.~70, no.~7, pp.~4900--4909, 2022.

\bibitem{FL_estimation_feedback} 
L.~Zhao, \emph{et al.}, ``Joint channel estimation and feedback for mm-wave system using federated learning,'' \emph{IEEE Commun. Lett.}, vol.~26, no.~8, pp.~1819--1823, 2022.

\bibitem{swin} 
Z.~Liu, \emph{et al.}, ``Swin Transformer: Hierarchical vision Transformer using shifted windows,'' in \emph{Proc. ICCV 2021}, Oct.~11-17, 2021, pp.~10012--10022.

\bibitem{kingma2013auto} 
D.~P.~Kingma and M.~Welling, ``Auto-encoding variational Bayes,'' \emph{arXiv preprint arXiv:1312.6114}, 2013.

\bibitem{fedams} 
Y.~Wang, L.~Lin, and J.~Chen, ``Communication-efficient adaptive federated learning,'' in \emph{Proc. ICML 2022} (Baltimore, MD, USA), Jul.~17-23, 2022, pp.~22802--22838.

\bibitem{xisuo_estiamtion_feedback} 
X.~Ma, Z.~Gao, F.~Gao, and M.~Di~Renzo, ``Model-driven deep learning based channel estimation and feedback for millimeter-wave massive hybrid MIMO systems,'' \emph{IEEE J. Sel. Areas Commun.}, vol.~39, no.~8, pp.~2388--2406, 2021.

\bibitem{bitcsinet} 
C.~Lu, W.~Xu, S.~Jin, and K.~Wang, ``Bit-level optimized neural network for multi-antenna channel quantization,'' \emph{IEEE Wireless Commun. Lett.}, vol.~9, no.~1, pp.~87--90, 2020.

\bibitem{torchsummary} 
T.~Yep, ``torch-summary 1.4.5,'' \url{https://pypi.org/project/torch-summary/}, accessed Dec. 24, 2020.

\end{thebibliography}
		
\end{spacing}
\end{document}